\documentclass[aps,preprint]{revtex4}%
\usepackage{amsfonts}
\usepackage{amsmath}
\usepackage{amssymb}
\usepackage{graphicx}%
\begin{document}
\title{{\LARGE THE PIONEER ANOMALY AND A MACHIAN UNIVERSE }}
\author{Marcelo Samuel Berman$^{1}$}
\affiliation{$^{1}$Instituto Albert Einstein / Latinamerica \ - Av. Candido Hartmann, 575 -
\ \# 17}
\affiliation{80730-440 - Curitiba - PR - Brazil \ msberman@alberteinsteininstitute.org}
\date{3 October, 2007}

\begin{abstract}
We discuss astronomical and astrophysical evidence, which we relate to the
principle of zero-total energy of the Universe, that imply several relations
among the mass \ $M$\ \ , the radius \ $R$\ \ \ and \ the angular momentum
\ $L$\ \ of a "large"\ sphere representing a Machian Universe. By calculating
the angular speed, we find a peculiar centripetal acceleration for the
Universe. This is an ubiquituous property that relates one observer to any
observable. It turns out that this is exactly the anomalous acceleration
observed on the Pioneers spaceships. We have thus, shown that this anomaly is
to be considered a property of the Machian Universe.\ \ We discuss several
possible arguments against our proposal.

\end{abstract}
\maketitle

\begin{center}
{\LARGE THE PIONEER ANOMALY AND A MACHIAN UNIVERSE }

\bigskip

Marcelo Samuel Berman
\end{center}

\bigskip\bigskip

{\Large I. Introduction}

\bigskip

In this paper, we discuss the so-called Pioneers' anomaly, which affects
spaceships sent to the outskirts of the Solar system, consisting of a
deceleration unaccounted by known physical causes.

\bigskip

We begin by describing Machian Universes, presenting \ this author's
viewpoint, which consists in defining such Universes by the condition
\ $E=0$\ , i.e., the total energy is zero, and time-invariant.\ We also
discuss some astrophysical similarities, to wit, the cosmological counterpart
of Blackett's and Wesson's laws.

\bigskip

We end this paper with a discussion on previous referees' considerations.

\bigskip

{\Large II. What is meant by a Machian Universe?}

Berman(2006b; 2007a; 2007b; 2007c), proposes that Mach's Principle, means a
zero-total energy Universe. Berman(2006; 2006a), has shown this meaning of
Mach's Principle without considering a rotating Universe. We now extend the
model, in order to include the spin of the Universe, \ and we replace
Brans-Dicke traditional relation, \ $\frac{GM}{c^{2}R}\sim1$\ \ , with two
different relations, which\ we call the Brans-Dicke relations for gravitation,
and for the spin of the Universe.

\bigskip

We shall consider a "large" sphere, with mass \ $M$\ \ , radius \ $R$\ \ ,
spin \ $L$\ \ .

\bigskip

We now calculate the total energy\ \ $E$\ \ of this distribution:

\bigskip

$E=E_{i}+E_{g}+E_{L}$\ \ \ \ \ \ \ \ \ \ \ \ \ \ , \ \ \ \ \ \ \ \ \ \ \ \ \ \ \ \ \ \ \ \ \ \ \ \ \ \ \ \ \ \ \ \ \ \ \ \ \ \ \ \ \ \ \ \ \ \ \ \ \ \ \ (1)

\bigskip

where \ \ $E_{i}=Mc^{2}$\ \ , stands for the inertial (Special
Relativistic)\ \ energy; \ $E_{g}\cong-\frac{GM^{2}}{R}$\ \ \ \ \ (the
Newtonian gravitational potential self-energy); \ \ $E_{L}\cong\frac{L^{2}%
}{MR^{2}}$\ \ the Newtonian rotational energy. Other contributions to the
total energy, might be added in relation (1), but we shall not do it here, for brevity.

\bigskip

If we impose that the total energy is equal to zero, i.e., $E=0$\ \ (Berman,
2006; 2006a; 2007a; 2007b; 2007c), we obtain from (1):

\bigskip

$\frac{GM}{c^{2}R}-\frac{L^{2}}{M^{2}c^{2}R^{2}}\cong1$%
\ \ \ \ \ \ \ \ \ \ \ \ \ \ . \ \ \ \ \ \ \ \ \ \ \ \ \ \ \ \ \ \ \ \ \ \ \ \ \ \ \ \ \ \ \ \ \ \ \ \ \ \ \ \ \ \ \ \ \ \ \ \ \ \ \ (2)

\bigskip

\bigskip As relation(2) above should be valid for the whole Universe, and not
only for \ a specific instant of time, in the life of the Universe, and if
this is not a coincidental relation, we can solve this equation by imposing
that \ $\dot{E}=0$\ \ (i.e., the zero-total-energy is a time-invariant
result), so that we are left with a single possible solution:

\bigskip

$\frac{GM}{c^{2}R}=\gamma_{G}$ \ \ \ \ \ \ \ \ \ \ \ \ \ \ \ \ \ \ \ \ \ \ \ , \ \ \ \ \ \ \ \ \ \ \ \ \ \ \ \ \ \ \ \ \ \ \ \ \ \ \ \ \ \ \ \ \ \ \ \ \ \ \ \ \ \ \ \ \ \ \ \ \ (3)

\bigskip

$\frac{L}{McR}=\gamma_{L}$\ \ \ \ \ \ \ \ \ \ \ \ \ \ \ \ \ \ \ \ \ \ \ , \ \ \ \ \ \ \ \ \ \ \ \ \ \ \ \ \ \ \ \ \ \ \ \ \ \ \ \ \ \ \ \ \ \ \ \ \ \ \ \ \ \ \ \ \ \ \ \ \ \ (4)

\bigskip

subject to the condition,

\bigskip

$\gamma_{G}-\gamma_{L}^{2}\cong1$\ \ \ \ \ \ \ \ \ \ \ \ \ \ \ \ \ , \ \ \ \ \ \ \ \ \ \ \ \ \ \ \ \ \ \ \ \ \ \ \ \ \ \ \ \ \ \ \ \ \ \ \ \ \ \ \ \ \ \ \ \ \ \ \ \ \ \ \ \ \ \ (5)

\bigskip

where the \ $\gamma^{\prime}s$\ \ are constants having a near unity value.

\bigskip

We now derive the following generalized Brans-Dicke relations, for gravitation
and spin:

\bigskip

$\frac{GM}{c^{2}R}=\gamma_{G}$ \ \ \ \ \ \ \ \ \ \ \ \ \ \ \ \ \ \ \ \ \ \ \ , \ \ \ \ \ \ \ \ \ \ \ \ \ \ \ \ \ \ \ \ \ \ \ \ \ \ \ \ \ \ \ \ \ \ \ \ \ \ \ \ \ \ \ \ \ \ \ \ \ (3)

\bigskip

$\frac{GL}{c^{3}R^{2}}=$ $\gamma_{L}$
\ \ \ \ \ \ \ \ \ \ \ \ \ \ \ \ \ \ \ \ \ \ \ , \ \ \ \ \ \ \ \ \ \ \ \ \ \ \ \ \ \ \ \ \ \ \ \ \ \ \ \ \ \ \ \ \ \ \ \ \ \ \ \ \ \ \ \ \ \ \ \ (6)

\bigskip

\bigskip Instead of deriving the above Brans-Dicke relations, by means of the
zero-total energy principle, coupled to the hypothesis that each term should
be valid, not only for the present Universe but also along all the history,
since Planck's Universe, other authors (Sabbata and Sivaram, 1994) derived the
B.D. relation for spin, on a heuristic procedure, which consists on the simple
hypothesis that \ $L$\ \ should obey a similar relation as $M$\ \ .\ 

\bigskip We notice that \ $R\propto M$\ \ \ , and \ $L\propto R^{2}$\ \ , in
case \ $\gamma_{G}$\ \ and \ $\gamma_{L}$\ \ are really constants.

\bigskip\ 

It must be remarked, that our proposed law (3), is a radical departure from
the original Brans-Dicke (Brans and Dicke, 1961) relation, which was an
approximate one, valid for the present Universe, while our present exact
hypotheses implies that \ \ $R\propto M$\ \ , and \ $L\propto R^{2}$\ , for
the entire span of the Universe's history.

\bigskip

With the present hypothesis, one can show, that independently of the
particular gravitational theory taken as valid, the energy density of the
Universe obeys a \ \ $R^{-2}$\ \ dependence (see Berman, 2006; 2006a; Berman
and Marinho, 2001). For instance, from the definition of the inertial or
matter energy density,

\bigskip

$\rho_{i}=\frac{M}{V}$\ \ \ \ \ \ \ \ \ , \ \ \ \ \ \ \ \ \ \ \ \ \ \ \ \ \ \ \ \ \ \ \ \ \ \ \ \ \ \ \ \ \ \ \ \ \ \ \ \ \ \ \ \ \ \ \ \ \ \ \ \ \ \ \ \ \ \ \ \ \ \ \ \ (7)

\bigskip

while, \ \ 

\bigskip

$V=\alpha R^{3}$\ \ \ \ \ \ ,\ \ \ \ \ ( $\alpha=$\ constant\ \ ) \ \ \ \ \ \ \ \ \ \ \ \ \ \ \ \ \ \ \ \ \ \ \ \ \ \ \ \ \ \ \ \ \ \ \ \ (8)

\bigskip

where \ $\rho_{i}$\ \ and \ \ $V$\ \ stand for energy density and
tridimensional volume, we find:

\bigskip

$\rho_{i}=\left[  \frac{\gamma_{G}}{G\text{ }\alpha}\right]  R^{-2}%
$\ \ \ \ \ \ \ \ \ . \ \ \ \ \ \ \ \ \ \ \ \ \ \ \ \ \ \ \ \ \ \ \ \ \ \ \ \ \ \ \ \ \ \ \ \ \ \ \ \ \ \ \ \ \ \ \ \ \ \ \ \ \ \ (9)

\bigskip

If we apply the above relation, for Planck's and the present Universe, we find:

\bigskip

$\frac{\rho_{i}}{\rho_{Pl}}=\left[  \frac{R}{R_{Pl}}\right]  ^{-2}$
\ \ \ \ \ \ \ \ \ \ \ . \ \ \ \ \ \ \ \ \ \ \ \ \ \ \ \ \ \ \ \ \ \ \ \ \ \ \ \ \ \ \ \ \ \ \ \ \ \ \ \ \ \ \ \ \ \ \ \ \ \ \ (10)

\bigskip

If we substitute the known values for Planck's quantities, while we take for
the present Universe, \ $R\cong10^{28}$\ cm, we find a reasonable result for
the present energy density. This shows that our result (relation 9), has to be
given credit.

\bigskip

\bigskip{\Large III. Pioneers' anomaly and the spin of the Universe}

It should be remembered that the origin of Planck's quantities, say, for
length, time, density and mass, were obtained by means of
dimensional\ \ combinations among the constants for macrophysics ($G$ for
gravitation and $c$ for electromagnetism) and for Quantum Physics (Planck's
constant $\frac{h}{2\pi}$). Analogously, if we would demand a dimensionally
correct Planck's spin, obviously we would find,

\bigskip

$L_{Pl}=\frac{h}{2\pi}$ \ \ \ \ \ \ \ \ \ \ \ \ \ \ . \ \ \ \ \ \ \ \ \ \ \ \ \ \ \ \ \ \ \ \ \ \ \ \ \ \ \ \ \ \ \ \ \ \ \ \ \ \ \ \ \ \ \ \ \ \ \ \ \ \ \ \ \ \ \ \ \ \ \ \ \ \ \ \ (11)

\bigskip

\bigskip This is exactly what we would obtain from (6), when we plug
\ $R_{Pl}$\ for \ $R$\ \ , and obtaining \ $L=L_{Pl}$\ \ \ .\ \ 

From Brans-Dicke relation for spin, we now can obtain the present angular
momentum of the Universe,

$\bigskip$

$L=L_{Pl}\left[  \frac{R}{R_{Pl}}\right]  ^{2}\cong10^{120}\left(  \frac
{h}{2\pi}\right)  =10^{93}$ $\ \ g$ $cm^{2}$ $s^{-1}$ \ \ \ \ \ \ \ . \ \ \ \ \ \ \ \ \ \ \ \ \ \ \ \ \ (12)

\bigskip

\bigskip This estimate was also made by Sabbata and Sivaram(1994), based on
heuristic considerations(see also Sabbata and Gasperini, 1979).

\bigskip

Sabbata and Gasperini(1979), have calculated the angular speed, for the
present Universe. Though they mixed their heuristic calculations with some
results obtained from Dirac's LNH (Large Number Hypothesis), including a time
variation for the gravitational "constant", we now show that, if we take for
granted that \ $G=$\ \ constant, and by means of the generalized Brans-Dicke
relations we find, by considering a rigid rotating Universe, whereby:

\bigskip

$L=MR^{2}\omega$\ \ \ \ \ \ \ \ \ \ , \ \ \ \ \ \ \ \ \ \ \ \ \ \ \ \ \ \ \ \ \ \ \ \ \ \ \ \ \ \ \ \ \ \ \ \ \ \ \ \ \ \ \ \ \ \ \ \ \ \ \ \ \ \ \ \ \ \ \ \ \ \ \ \ \ \ \ (13)

\bigskip

so that,

\bigskip

$M\omega=$\ constant\ \ \ , ( because \ $L\propto R^{2}$\ \ \ as we have shown
earlier ), we shall have:

\bigskip

$\omega_{Pl}^{{}}=\frac{c}{R_{Pl}}=2\ $x $10^{43}$ \ $s^{-1}$ \ \ \ \ \ , \ \ \ \ \ \ \ \ \ \ \ \ \ \ \ \ \ \ \ \ \ \ \ \ \ \ \ \ \ \ \ \ \ \ \ \ \ \ \ \ \ \ \ \ \ \ \ \ (14)

\bigskip

and, for the present,

\bigskip

$\omega=\frac{c}{R}\cong3$ x $10^{-18}$ \ $s^{-1}$
\ \ \ \ \ \ \ \ \ \ \ \ \ \ . \ \ \ \ \ \ \ \ \ \ \ \ \ \ \ \ \ \ \ \ \ \ \ \ \ \ \ \ \ \ \ \ \ \ \ \ \ \ \ \ \ (15)

\bigskip

\bigskip Sabbata and Gasperini(1979), pointed out that the same numerical
angular speed is obtained for G\"{o}del's Universe, and also for the Sun's
peculiar velocity through the cosmic microwave background.

\bigskip

We remark that \ $\gamma_{G}\cong2$\ \ is to be exact and not approximate, if
we consider the result by Adler et al (1975), for the energy of a spherical
mass, obtained by means of pseudotensors.

\bigskip

The Pioneers' anomaly, \ is described by a centripetal acceleration of an up
to now unexplained nature, which affects two spaceships launched on opposite
directions, which are by now in the outskirts of the Solar system (Anderson,
1999). Its value is \ $a^{\prime}\cong-8$\ x\ $10^{-8}cm/\sec^{2}$\ .\ \ 

\bigskip

For a Machian Universe, taken care of result (15),we can obtain the value for
an ubiquitous centripetal acceleration,

\bigskip

$a=-\omega^{2}R$ \ \ \ \ \ \ \ \ \ \ \ \ \ \ \ \ \ \ \ \ . \ \ \ \ \ \ \ \ \ \ \ \ \ \ \ \ \ \ \ \ \ \ \ \ \ \ \ \ \ \ \ \ \ \ \ \ \ \ \ \ \ \ \ \ \ \ \ \ \ \ \ \ \ \ \ \ (16)

\bigskip

If \ \ $R\cong10^{28}cm$\ \ , \ \ as is known for the causally related
Universe, we find:

\bigskip

$a=-9$\ x\ $10^{-8}cm/\sec^{2}\cong a^{\prime}$\ \ \ \ \ \ .\ \ \ \ \ \ \ \ \ \ \ \ \ \ \ \ \ \ \ \ \ \ \ \ \ \ \ \ \ \ \ \ \ \ \ \ \ \ \ \ \ \ \ \ \ \ (17)

\bigskip It is necessary to point out that, for a Machian Universe, we should
have this extra acceleration, along the direction pointing from the observed
to the observer. It affects any two pairs of, observer versus observed, points
in space. The striking match between \ \ $a$\ \ and \ \ $a^{\prime}$ \ \ must
point to a possible solution to the Pioneers' anomaly; the only necessary
hypothesis is that the Universe is endowed with the Machian properties shown above.

\bigskip

\bigskip{\Large IV. Astrophysical and Cosmological Laws}

\bigskip There are two astrophysical empirical laws, called after Blackett,
and Wesson, relating, the first, spins and magnetic moments of astrophysical
objects and stars; the second, relating spins and masses, of the same
objects(Sabbata and Sivaram, 1994; Wesson, 2006).

\bigskip

If we call by\ \ $U_{a}$\ \ , the magnetic moment, it is found the approximate
relation for astrophysical spins \ \ $L_{a}$\ \ \ ,

\bigskip

$L_{a}=qU_{a}$\ \ \ \ \ \ \ ( \ $q\approx10^{15}$\ \ \ \ \ g$^{1/2}%
$.cm$^{-1/2}$ \ ) \ \ \ \ \ \ \ \ \ \ \ \ \ \ \ \ \ \ ,

\bigskip

while,

\bigskip

$L_{a}=pM_{a}^{2}$ \ \ \ \ \ \ \ \ ( \ \ $p\approx10^{-15}$\ \ \ \ g$^{-1}%
$.cm$^{2}$sec$^{-1}$\ ) \ \ \ \ \ \ \ \ \ \ \ \ \ \ \ \ \ \ ,

\bigskip

where \ \ $M_{a}$\ \ \ represent the masses of the objects.

\bigskip

If we remember the Machian properties of Sections II and III, we find that the
last relation is obeyed by the Machian Universe, with a not very smaller value
for the constant \ $p$\ \ , say\ \ \ $10^{-17}$\ \ \ \ \ g$^{-1}$.cm$^{2}%
$sec$^{-1}$\ \ . The astrophysical law is verified for several different
objects; the Universe, obeys such law, at any instant, and we guess that the
astrophysical value of \ \ $p$\ \ , will approach the Universe's one, as much
as the observed objects have larger masses.

\bigskip

As to Blackett's law, we find that the Machian Universe should also obey it,
with a not too much different numerical value for the constant \ $q$\ , and we
also guess that the larger \ the mass of the objects, the more, the numerical
values of the objects \ $p$\ \ \ will approach the one for the Universe.

\bigskip

\bigskip It must be remembered, that the magnetic field of the Universe must
obey the same \ \ $R^{-2}$\ -- dependence for its energy density, i.e.,

\bigskip

$\frac{B^{2}}{8\pi}\propto R^{-2}$ \ \ \ \ \ \ \ \ \ \ \ \ \ .

\bigskip

We shall also need to add one more term in equation \ (1), in order to
represent the magnetic field's energy\ contribution,

\bigskip

$E_{B}=\frac{B^{2}}{8\pi}\cdot\frac{4\pi R^{3}}{3}$ \ \ \ \ \ \ \ \ \ \ \ \ \ \ .

\bigskip

From experimental arguments, we fix the present value for \ $B$\ \ \ to be of
order \ \ $10^{-6}$\ \ Gauss.

\bigskip

Hence, the approximate numerical value for the Machian's \ $q$\ \ .

\bigskip

\bigskip{\Large V. Pros and Cons of our Machian picture}

\bigskip We may argue that (1) it would be unclear who should measure the
energy of the Universe, from the "outside"; (2) it would be unclear whether we
may use Newtonian expressions for the calculations; (3) it would be
mathematically impossible to derive several generalized Brans-Dicke
equalities, from a single equation describing the energy \ $E$\ ; \ (4) the
local energy-momentum conservation, described by the covariant divergence of
the energy-momentum tensor, would be no more valid, and therefore, the model
is inconsistent; (5) the large angular-momentum of the Universe, is not
astronomically confirmed; (6) this paper does not obey any viable theory of
Gravity, and it does not supply new results about the Universe; (7) the
Brans-Dicke relation is numerically verified for the present Universe, but the
generalized counterpart, which is an equality, is obviously also verified, so
that, nothing new has been provided, and, the coincidence has a lot of
uncertainty; (8) what Berman is doing, is just an exercise in dimensional
analysis, like has been earlier done for instance, by Dirac and Eddington; (9)
this theory is heuristic, and, thus, not necessarily scientific.

\bigskip

However, we answer those "cons", with the following "pros": (A) allegations
about the energy of the Universe, and, precisely, about its zero-value, can be
traced to Feynman (Feynman, 1962-3), Rosen (Rosen, 1994-95), Cooperstock and
Israelit (1995), Hawking (2001) and many others. Berman has derived this from
Robertson-Walker's metric, so that it is a valid result in Relativistic
Cosmology, for any tri-curvature value (Berman, 2006, 2006a). The existence of
a "spectator" is a philosophical question, rather than a scientific one; (B)
Machian properties have been proposed in different gravity theories, so there
is no one single theory that owns such attributes (remember the origin of
Brans-Dicke theory); (C) the several generalized Brans-Dicke equalities,
derived from the energy equation, are just, the most simple set of solutions
for the \ $E=0$\ \ equation; (D) the mentioned solutions, have very
interesting properties: for instance, the relative contributions of each type
of energy towards the total amount, is time-independent. This fact is coherent
with the recently proclaimed and experimentally observed result that the
Universe has been \ \ lambda-dominated since long ago; (E)\ we never told that
"Machian" conditions only can mean "general relativistic" ones; (F) you can
not blame our paper for the fact that the angular momentum is high for the
present Universe, because we have derived a correct result, i.e., the small
amount of angular velocity in the present Universe, which angular velocity is
undetectable with present technological tools; (G) our framework is
relativistic, in the low Newtonian limit, but this could be called, also, a
Sciama gravitational theory (Sciama, 1953); (H) we can extend all forms of
energy densities towards Planck's time, by going back from the present: no
inconsistency with Planck's energy density would be found. It must be not
overlooked that the effective energy density of the Universe is zero-valued,
corresponding to a zero-total energy. This is attained by subtracting, from
all kinds of energy densities (corresponding to inertial mass, spin,
\ cosmological constant, radiation, etc.), \ which are positive, the energy
density due to the self-gravitational term, which is negative, and balances
the first ones.

\bigskip

We refer to the extremely important books by Sabbata and Sivaram(1994) and
Wesson(2006), where there are clues about the rotation of the Universe, for
instance, through Blackett and Wesson's formulae, which relates spin and
magnetic field.

\bigskip

{\Large Acknowledgements}

\bigskip The author gratefully thanks his intellectual mentors, Fernando de
Mello Gomide and M. M. Som, his colleagues Nelson Suga and Mauro Tonasse, and
Marcelo F. Guimar\~{a}es; I am also grateful for the encouragement by Paula,
Albert and Geni. The last referee, was also very helpful, by arising a
question on the solution of equation (2), \ which indeed depends on the
assumption \ $\dot{E}=E=0$\ \ .

\bigskip

{\Large References}

\bigskip

Adler, R.J.; Bazin, M.; and Schiffer, M. (1975) - \textit{Introduction to
General Relativity} , McGraw-Hill, 2nd. Ed., New York.

Anderson, J.D. (1999) - \textit{Planetary Report}, \textbf{19(3)}, 15.

\bigskip Berman,M.S. (2006) - \textit{Energy of Black-Holes and Hawking's
Universe \ }in \textit{Trends in Black-Hole Research, }Chapter 5\textit{.}
Edited by Paul Kreitler, Nova Science, New York.

Berman,M.S. (2006 a) - \textit{Energy, Brief History of Black-Holes, and
Hawking's Universe }in \textit{New Developments in Black-Hole Research},
Chapter 5\textit{.} Edited by Paul Kreitler, Nova Science, New York.

Berman,M.S. (2006 b) - \textit{On the Machian Properties of the Universe}, submitted.

Berman,M.S. (2007 a) - \textit{Introduction to General Relativity and the
Cosmological Constant Problem}, Nova Science, New York.

Berman,M.S. (2007 b) - \textit{Is the Universe a White-Hole?}, Astrophysics
and Space Science, at press. See Los Alamos Archives,
http://arxiv.org/abs/physics/0612007 .

Berman,M.S. (2007 c) - \textit{Introduction to General Relativistic and
Scalar-Tensor Cosmologies}, Nova Science, New York.

Berman,M.S.; Marinho,R.M. (2001) - Astrophysics and Space Science,
\textbf{278}, 367.

Brans, C.; Dicke, R.H. (1961) - Physical Review, \textbf{124}, 925.

Cooperstock, F.I.; Israelit, M. (1995) - Foundations of Physics, \textbf{25}, 631.

Feynman, R. (1962-3) - \textit{Lectures on Gravitation}, Addison-Wesley, N.Y.

Hawking, S.W. (2001) - \textit{The Universe in a Nutshell, }Bantam, N.Y.

Rosen, N. (1994) - GRG \textbf{26}, 319.

\bigskip Sabbata, V.; Sivaram, C. (1994) - \textit{Spin and Torsion in
Gravitation,} World Scientific, Singapore.

Sabbata, V.; Gasperini,M.\ (1979) - Lettere Nuovo Cimento \textbf{25}, 489.

\bigskip Sciama, D.N. (1953) - MNRAS \textbf{113}, 34.

\bigskip Wesson, P.S. (2006) - \textit{Five Dimensional Physics}, World
Scientific, Singapore.

\end{document}